\documentclass[reprint,amsmath,amssymb,aps,pra,superscriptaddress,floatfix,nofootinbib]{revtex4-2}

\usepackage[utf8]{inputenc}
\usepackage[english]{babel}
\usepackage{hyperref}
\usepackage{bm}
\usepackage{hyperref}
\usepackage{physics}
\usepackage[dvipsnames]{xcolor}
\usepackage{booktabs}
\usepackage{float}
\usepackage{graphicx}
\usepackage[capitalise]{cleveref}
\usepackage{mhchem}
\usepackage{tabularx}
\usepackage{ulem}
\usepackage{enumitem}
\usepackage[
    protrusion=true,
    activate={true,nocompatibility},
    final,
    tracking=true,
    kerning=true,
    spacing=true,
    factor=1000]{microtype}

\hypersetup{
 colorlinks=true,       
 linkcolor=blue,          
 citecolor=blue,        
 filecolor=blue,      
 urlcolor=blue}

\newcommand{\beginsupplement}{%
    \setcounter{table}{0}
    \setcounter{figure}{0}
    \setcounter{equation}{0}
    \renewcommand{\thetable}{S\arabic{table}}%
    \renewcommand{\thefigure}{S\arabic{figure}}%
    \renewcommand{\thesection}{S-\arabic{section}}
    \renewcommand{\theequation}{S.\arabic{equation}}
    \widetext
}
\usepackage{titlesec}
\titleformat*{\section}{\bfseries\Large}
\titleformat*{\subsection}{\bfseries}
\usepackage{caption}

\newcommand{\mr}[1]{\mathrm{#1}}

\begin{document}

                                                         
\author{Gr\'egory Moille}
\email{gmoille@umd.edu}
\affiliation{Joint Quantum Institute, NIST/University of Maryland, College Park, USA}
\affiliation{Microsystems and Nanotechnology Division, National Institute of Standards and Technology, Gaithersburg, USA}
\author{Jordan Stone}
\affiliation{Joint Quantum Institute, NIST/University of Maryland, College Park, USA}
\affiliation{Microsystems and Nanotechnology Division, National Institute of Standards and Technology, Gaithersburg, USA}
\author{Michal Chojnacky}
\affiliation{Joint Quantum Institute, NIST/University of Maryland, College Park, USA}
\affiliation{Sensor Science Division,, National Institute of Standards and Technology, Gaithersburg, USA}
\author{Curtis Menyuk}
\affiliation{University of Maryland at Baltimore County, Baltimore, MD, USA}
\author{Kartik Srinivasan}
\affiliation{Joint Quantum Institute, NIST/University of Maryland, College Park, USA}
\affiliation{Microsystems and Nanotechnology Division, National Institute of Standards and Technology, Gaithersburg, USA}
\date{\today}


\title{Kerr-Induced Synchronization of a Cavity Soliton to an Optical Reference for Integrated Frequency Comb Clockworks}

                                                             
\begin{abstract}
    \noindent The phase-coherent frequency division of a stabilized optical reference laser to the microwave domain is made possible by optical frequency combs (OFCs). Fundamentally, OFC-based clockworks rely on the ability to lock one comb tooth to this reference laser, which probes a stable atomic transition. The active feedback process associated with locking the comb tooth to the reference laser introduces complexity, bandwidth, and power requirements that, in the context of chip-scale technologies, complicate the push to fully integrate OFC photonics and electronics for fieldable clock applications. Here, we demonstrate passive, electronics-free synchronization of a microresonator-based dissipative Kerr soliton (DKS) OFC to a reference laser. We show that the Kerr nonlinearity within the same resonator in which the DKS is generated enables phase locking of the DKS to the externally injected reference. We present a theoretical model to explain this Kerr-induced synchronization (KIS), and find that its predictions for the conditions under which synchronization occur closely match experiments based on a chip-integrated, silicon nitride microring resonator. Once synchronized, the reference laser is effectively an OFC tooth, which we show, theoretically and experimentally, enables through its frequency tuning the direct external control of the OFC repetition rate. Finally, we examine the short- and long-term stability of the DKS repetition rate and show that the repetition rate stability is consistent with the frequency division of the expected optical clockwork system.
 \end{abstract}
\maketitle


                          
Optical frequency combs (OFCs) are unique metrological tools that provide a phase-coherent link between the optical and microwave domains, and serve as the basis for many metrology applications, such as optical synthesizers~\cite{ma_optical_2004,SpencerNature2018} and distance ranging~\cite{coddington_rapid_2009,RiemensbergerNature2020}. %
Furthermore, OFCs are essential components in time-keeping applications~\cite{DiddamsScience2020a}. %
Although a microwave frequency Cs atomic clock continues to be the primary standard that defines the second~\cite{TiesingaRev.Mod.Phys.2021}, the ability to improve clock performance has led to the use of atoms having energy transitions in the optical spectrum~\cite{LudlowRev.Mod.Phys.2015}. %
For optical atomic clocks, a stable laser, referred to as the reference laser in this work, is locked to this optical frequency transition~\cite{diddams_optical_2001,MartinPhys.Rev.Appl.2018, BothwellMetrologia2019}. %
Direct electronic measurement of the locked laser frequency is usually not possible, so an OFC is employed as an optical-to-microwave frequency divider~\cite{FortierCommunPhys2019}. %
The clock output can be determined by measuring the OFC repetition rate directly, which carries the reference laser stability~\cite{holzwarth_optical_2001, PappOpticaOPTICA2014b}. %
Recently, the advent of integrated photonics has made optical clockwork utilizing an OFC available on-chip~\cite{NewmanOptica2019,DrakePhys.Rev.X2019}. %
The integrated photonics OFC is typically realized in a microring resonator through the creation of a dissipative Kerr soliton (DKS), in which the material's third-order nonlinearity is employed to convert a continuous wave (CW) optical pump into a single solitary pulse traveling throughout the cavity~\cite{KippenbergScience2018}. %
The periodic extraction of this soliton forms a pulse train in the bus waveguide, which yields the OFC via Fourier relationship. %
Recently, octave-spanning on-chip OFCs~\cite{li_stably_2017, OkawachiOpt.Lett.OL2011a, PfeifferOpticaOPTICA2017} whose spectral extent reaches atomic transitions~\cite{YuPhys.Rev.Applied2019a} have been experimentally demonstrated, enabling a direct interface with the reference laser. %
Whether the OFCs are on a chip or not, the clockwork for optical frequency division using them entails the active locking of a single OFC comb tooth to the reference laser, where the beat note is recorded and stabilized using an active feedback loop to the OFC external degrees of freedom (often its pump current and frequency). %
Though there has been much progress in developing electronics associated with OFC applications~\cite{ManurkarOSAContinuumOSAC2018, ShawOSAContinuumOSAC2019}, the power budget and limited bandwidth associated with such active stabilization represents a significant barrier to full integration of an optical clock, even if the power required to optically drive the integrated OFC can be extremely low~\cite{SternNature2018b}. %
\begin{figure*}[t]
    \begin{center}
        \includegraphics[width = \textwidth]{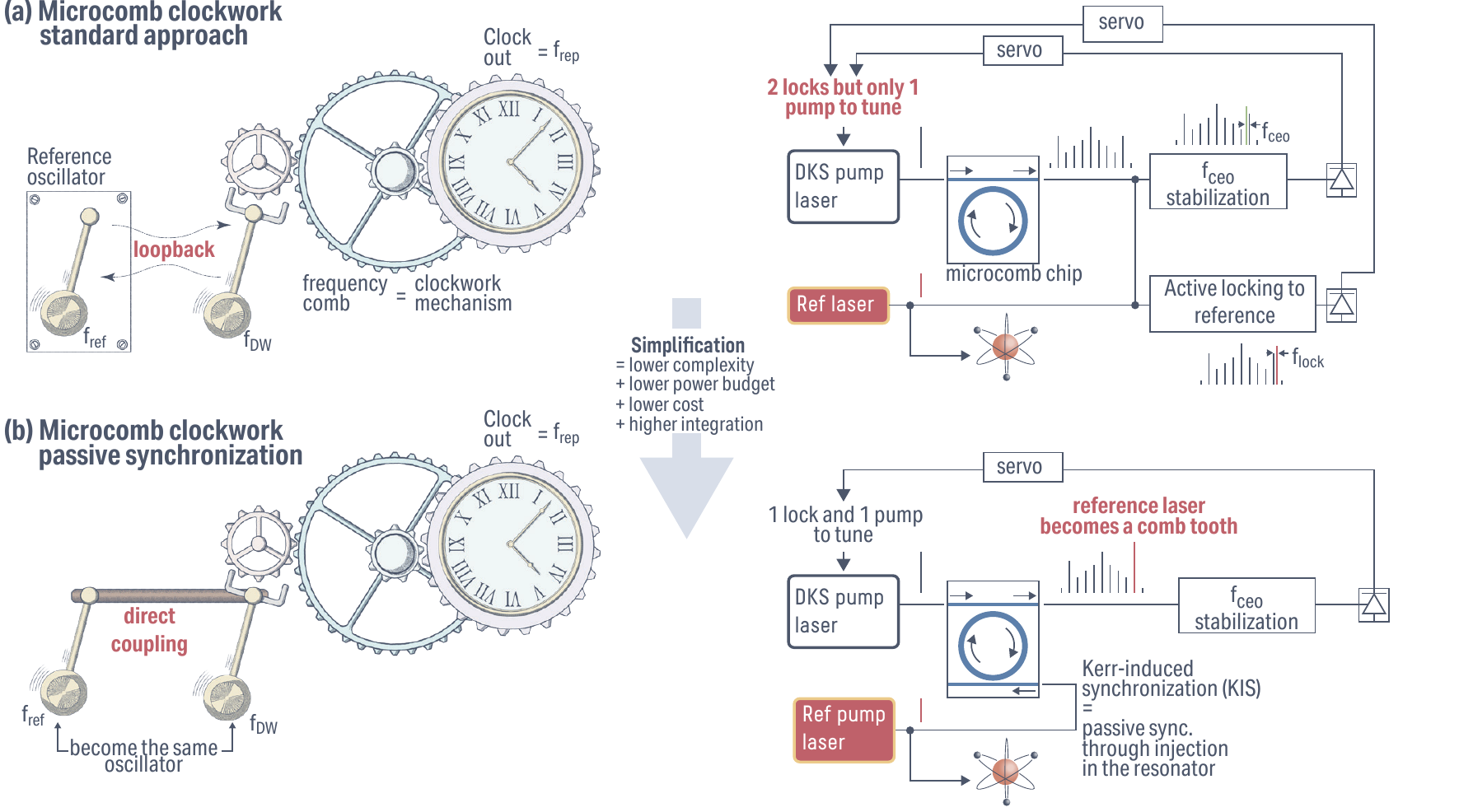}
    \end{center}
    \caption{\label{fig:1} \textbf{Clockwork concept} -- %
    \textbf{(a)} Mechanical Shortt's clock concept (left) and equivalent optical clockwork. This scheme corresponds to the current approach to link a reference laser, whose frequency is stabilized to an atomic transition $f_\mr{ref}$, to one of the microcomb comb teeth (usually at the dispersive wave $f_\mr{DW}$), for optical frequency division down to the microcomb repetition rate $f_\mr{rep}$. The active feedback for locking the comb to the reference introduces additional power, complexity, and bandwidth considerations, resulting in significant challenges for full chip-scale integration of an optical clock. %
    \textbf{(b)} Mechanical Huygens' clock concept (left) and equivalent optical clockwork. This scheme corresponds to the concept of passive Kerr-induced synchronization (KIS) between the reference and the comb, which we study in this work. In the mechanical clock, the coupling between the two pendulums is lossy and results in poor long term stability. In the optical regime, the coupling can be provided by the lossless optical Kerr nonlinearity. The coupling link between the DKS and the reference laser created by injecting the latter directly into the DKS microresonator, where it is another pump for the circulating DKS and enables passive synchronization for a significantly simplified optical clockwork. %
   }
\end{figure*}

In this work, we show that injecting the reference laser into a DKS microresonator -- hence turning it into another pump for the DKS system -- allows for all-optical, passive Kerr-induced synchronization (KIS) between the OFC and the reference laser. %
Nonlinear Kerr phase locking of the DKS to the reference pump laser is demonstrated by using a microresonator with the appropriate dispersion for exhibiting a dispersive wave at the reference pump frequency. %
We investigate the system's nonlinear dynamics, where the experimental findings exhibit excellent agreement with the Lugiato-Lefever equation. %
Additionally, we derive an expanded Adler equation that describes the requirements for synchronization and its behavior. When synchronized, the reference pump is effectively part of the DKS, and we accordingly show that the DKS repetition rate may be adjusted on demand by detuning the reference pump, with the resulting repetition rate tuning in accordance with a frequency division ratio set by the separation between the reference pump and the DKS-generating laser. %
Finally, we demonstrate that this passive Kerr-induced synchronization is compatible with clock operation by showing both division of the phase stability of the reference laser onto the OFC repetition rate and the division of the long-term stability.

The analogy between optical and mechanical clockworks is frequently used~\cite{DiddamsScience2020a}, with the OFC portrayed as a set of gears [\cref{fig:1}]. %
To better understand the context of our work, it is compelling to consider the development of the mechanical pendulum clock, as exemplified by Shortt's implementation~\cite{Riehle_frequency_standards_book}, which achieved the highest accuracy for clocks in its era. %
By actively locking a follower pendulum to a reference vacuum-sealed pendulum not connected to a clockwork, the reference avoids mechanical damping, which is the primary factor contributing to the poor long-term stability of the pendulum [\cref{fig:1}a]. %
Essentially, the same architecture as Shortt's is replicated by contemporary OFC clockworks. %
One can consider synchronization as an alternative to active locking, as Huygens famously established that direct synchronization between two pendulums could happen if a straightforward coupling mechanism is offered [\cref{fig:1}(b)]. %
However, this mechanical architecture offers poor long-term stability because of the coupling's increased clock damping. %
While the similarities to the history of mechanical clockworks are compelling, an important distinction with respect to optical clocks is in the damping mechanisms. %
In particular, optical nonlinearities, which are mostly lossless, can be used to produce the coupling necessary for long-term synchronization. %
Indeed, such nonlinear synchronization has been observed, for example in counter-propagating DKSs in a microresonator~\cite{YangNaturePhoton2017}. %
Therefore, in analogy with a mechanical clockwork, designing an OFC architecture that passively synchronizes the reference laser and the OFC through Kerr-induced nonlinear coupling should be possible. %
This would significantly reduce optical clock complexity, size, and power requirements. 

\begin{figure*}[t]
    \begin{center}
        \includegraphics{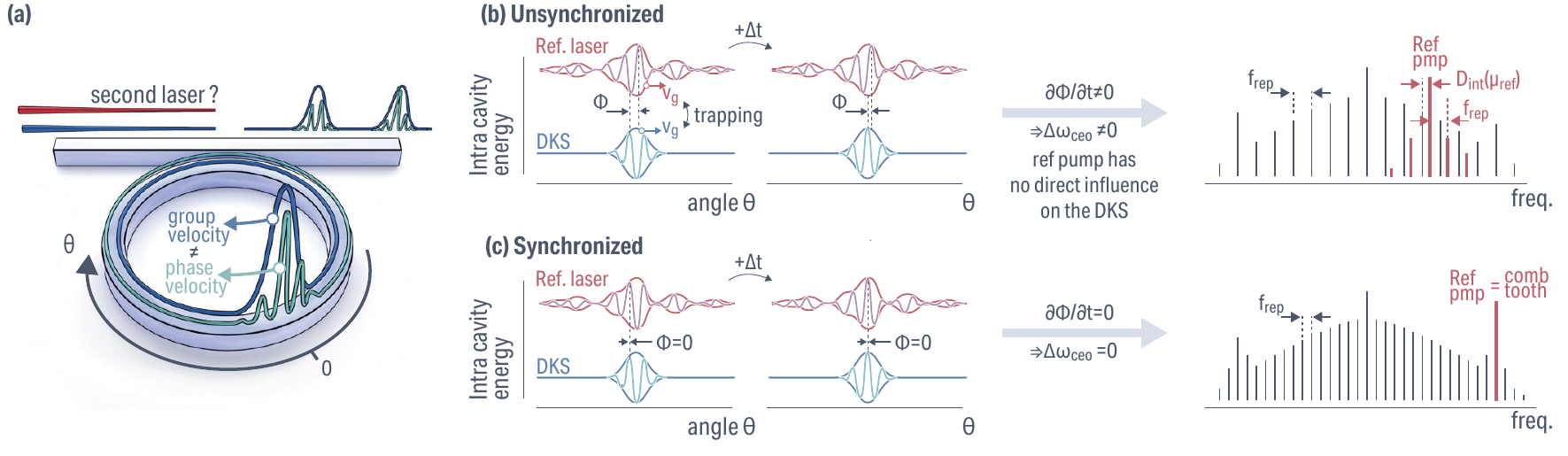}
    \end{center}
    \caption{\label{fig:2} \textbf{Concept of phase locking for Kerr-induced synchronization (KIS)} -- %
    \textbf{(a)} DKS circulating in a resonator where the envelope travels at the group velocity while the fast oscillations travel at a distinct phase velocity, resulting in a non-zero carrier envelope offset. A second, reference pump laser can also be injected into the DKS resonator. %
    \textbf{(b)} In the unsynchronized case, the phase velocities of the reference and pump components of the MDKS are different, yielding a constant phase slip in time, and hence an offset in CEO observable in the comb spectrum. %
    \textbf{(c)} In the KIS case, there is no temporal phase slip between the components, and thus the MDKS becomes a single DKS. Consequently, no CEO frequency offset is observable as the reference pump has become a comb tooth.%
    }
\end{figure*}
\vspace{1ex}
Before delving into the experimental findings, it is necessary to recall the DKS behavior under various driving conditions. %
The group and phase velocity are the two fundamental characteristics of the DKS. %
The repetition rate $\omega_\mr{rep}$ arises from the former, and specifies the temporal separation between two extracted pulses in the bus waveguide. %
The DKS phase velocity is connected to the carrier-envelope offset (CEO), which is a phase slip $\varphi(t)$ of the rapid oscillation below the envelope for each extracted pulse, and results in the CEO frequency $\omega_\mr{ceo} = \partial \varphi/\partial t $ [\cref{fig:2}(a)]. %
When the reference pump laser is injected in the cavity, a secondary pulse is created that is locked to the DKS by cross phase modulation (XPM), so that a multi-color soliton (MDKS) is created that moves at a single group velocity in the microresonator~\cite{WangOptica2017, MoillearXiv2023}. As a result of the single group velocity MDKS, any comb formed around the reference pump will have the same $\omega_\mr{rep}$ as the DKS OFC. %
To accomplish effective power transfer inside the cavity, the reference must be on resonance when it is injected into the microring. %
However, the two components of the MDKS will in general have different phase velocities that are determined by cavity dispersion, leading to a non-zero CEO frequency shift between these two comb components $\Omega = \omega_\mr{ref} - \omega_\mr{\mu,s}=\partial \Phi /\partial t $ [\cref{fig:2}(b) - top], where $\Phi = \varphi_\mr{ref} - \varphi_\mr{dks}$, and $\omega_\mr{\mu,s}$ is the frequency at mode $\mu_s$ (the comb tooth closest to the reference pump frequency). %
It has been demonstrated that this phase velocity mismatch can be utilized to produce coherent optical parametric oscillation in the phase velocity domain, yielding an ultra-broadband comb through spectral extension~\cite{MoilleNat.Commun.2021a} or even cascading to form a so-called two-dimensional comb~\cite{MoillearXiv2023}, where each comb component exhibits the same repetition rate with a constant frequency offset between them [\cref{fig:2}(b)]. %
These demonstrations are intriguing, but by definition, systems exhibiting $\Omega = \partial \Phi /\partial t \neq 0$ are not synchronized. %
As a result, the OFC is not phase synchronized to the reference pump, and such a system cannot be used as a coherent clockwork. %
When the reference and the DKS are synchronized, $\partial \Phi /\partial t =0 $ [\cref{fig:2}(c)], and no frequency offset is observable. %
In this situation, it becomes impossible to distinguish between one of the DKS comb teeth and the reference pump as the latter effectively becomes a comb tooth. %
Until now, multi-driving of a DKS has only been observed when the primary and secondary combs are unsynchronized~\cite{MoilleNat.Commun.2021a, MoillearXiv2023, QureshiCommunPhys2022, ZhangNatCommun2020} and when soliton crystals are generated~\cite{LuNatCommun2021, TaheriNatCommun2022a}. %
Here, we investigate the specific conditions to achieve phase locking and demonstrate its appearance. \\
\begin{figure*}[t]
    \begin{center}
        \includegraphics{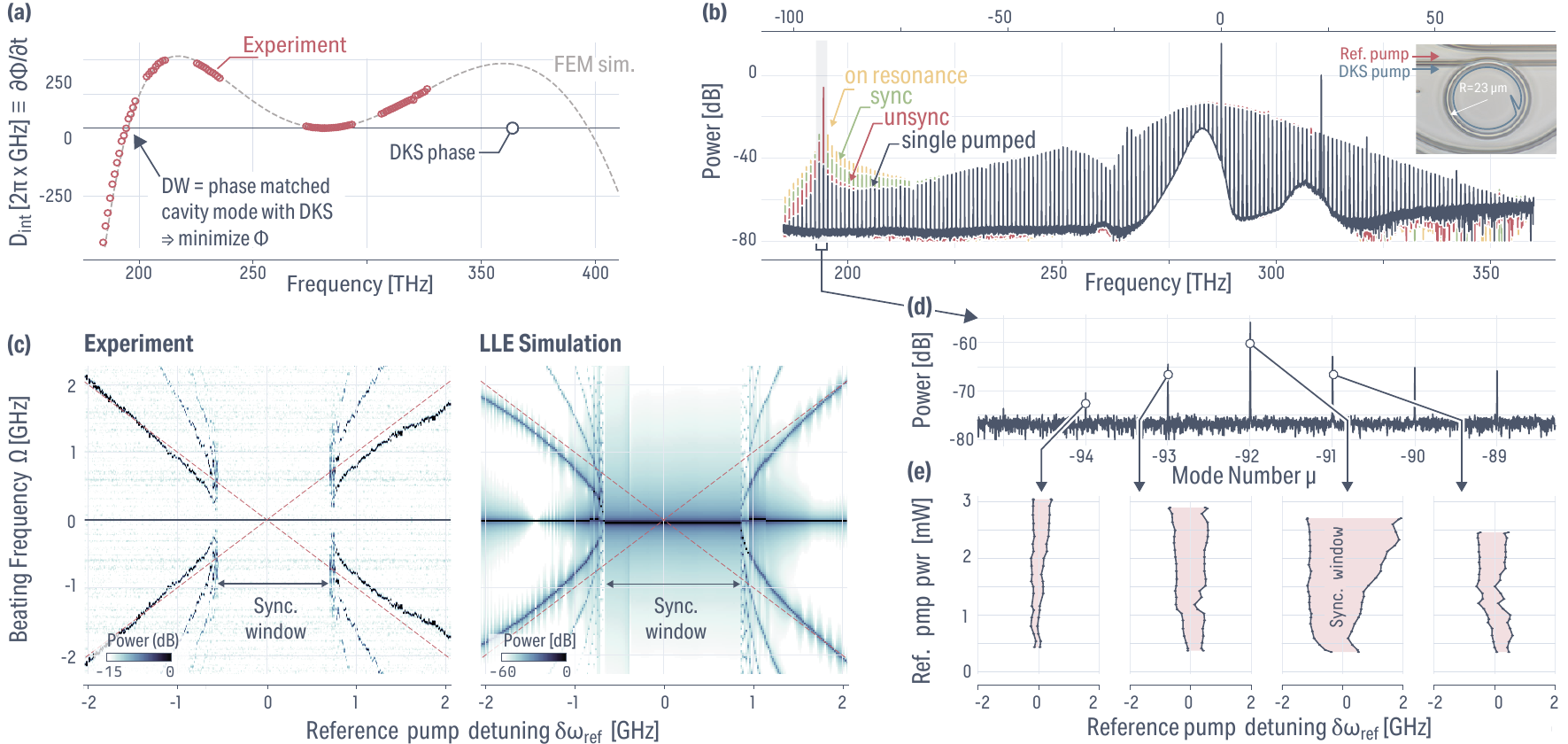}
    \end{center}
    \caption{\label{fig:3}%
    \textbf{Experimental demonstration of synchronization.} -- %
    \textbf{(a)} Experimental (open circles) and simulated (dashed line) integrated dispersion of the microring resonator under study, presenting an accessible dispersive wave (DW) at $\mu \approx -92$ ($\approx194$~THz), where injection of the reference pump into the cavity at a frequency near that of a DKS comb tooth is possible. The other (high frequency) DW is not experimentally accessible because of the pulley-coupling that has been chosen to optimize pump and reference coupling to the microring. %
    \textbf{(b)} OFC obtained while pumping at 284~THz with a cooler at 308~THz for thermal stability, and a reference pump near 194~THz. The different regimes of the DKS with different detunings of the reference at $\mu = -91$ are highlighted. %
    \textbf{(c)} Recording of the CEO frequency mismatch $\Omega$ ($y$ axis) with the reference pump detuning $\delta\omega_\mr{ref}$ ($x$ axis). The absence of a beat in the central region is a signature of the synchronization. The experimental data (left) are reproduced accurately by the LLE (right). %
    \textbf{(d)} Zoom-in around the DW mode, exhbiting lower comb tooth power the further from $\mu=-92$ the azimuthal component is.  %
    \textbf{(e)} Frequency bandwidth of the synchronization window with the reference pump power and azimuthal mode number into which the reference pump is injected. As expected by the Adler equation, the synchronization window bandwidth decreases as both the reference pump power and the azimuthal DKS component power decrease. %
    }
\end{figure*}
\indent Synchronization between oscillators can be understood using an Adler equation, for which an extended version can be obtained for a DKS under multiple driving fields (see supplementary material~\cref{sup:adler}), such that: 
\begin{equation}
 \label{eq:adler}
 - \frac{1}{\kappa} \frac{\partial^2\Phi}{\partial{t}^2} - \frac{\partial\Phi}{\partial{t}} =D_\mathrm{int}(\mu_s) + \Delta - \mathcal{T} \sin{\left(\Phi \right)} 
\end{equation}%
with $\mu$ the azimuthal order referenced to the main pump, $\Delta = \delta\omega_\mr{ref} + \mu_sD_1(\mu_0)(1-K_\mathrm{NL} - K_0)$ the effective reference pump detuning with $\delta \omega_\mr{ref}$ being the detuning of the reference pump, $K_0 =\frac{2 \left|{A(\mu_0)}\right|}{E_\mr{DKS}} \sqrt{\frac{P_\mr{main} \kappa_\mr{ext}}{\kappa^{2}}}$ the normalized XPM phase shift induced by the mode $\mu_\mr{j}$, $P_\mr{ref}$ and $E_\mr{dks}$ are the power and energy of the reference and the DKS respectively, $D_1(\mu_\mr{j})$ is the linear free spectral range at the azimuthal component $\mu_\mr{j}$, $K_\mr{NL}$ accounts for the cross-phase modulation induced by the MDKS (see \cref{sup:adler}), $\kappa$ and $\kappa_\mr{ext}$ are the total and coupling losses, respectively, $\mu_s$ is the azimuthal mode at which the reference is synchronized, and $\mathcal{T}=D_{1}(\mu_{s}) \mu_{s} \frac{2 \left|{A(\mu_{s})}\right|}{E_\mr{DKS}} \sqrt{\frac{P_\mr{ref} \kappa_\mr{ext}}{\kappa^{2}}}$ is the effective natural torque of the system. %
\cref{eq:adler} is similar to the equation that is obtained for counter-propagating DKSs~\cite{YangNaturePhoton2017}, though in our work the synchronization is provided by an external driving force rather than another intra-cavity state. %
The synchronization state occurs when the left-hand side of \cref{eq:adler} is null, and hence $D_\mr{int}(\mu_s)$ must be compensated either by the nonlinear shift or the reference pump detuning. Although it is theoretically possible to achieve this with a large CEO offset from $D_\mr{int}(\mu_s)$ that can be compensated by XPM, the complex balance of the phase shift at $\mu_s$ ($K_{\mu, s}$) and the high power involved would render experimental demonstration challenging. %
Therefore, minimizing $D_\mr{int}(\mu_s)$ allows minimization of the reference power for which the synchronization effect still happens.\\
\indent It has been demonstrated that integrated microring resonators can produce an octave-spanning comb, partly due to the generation of dispersive waves (DWs) produced by phase-matching cavity resonances with the DKS. %
High order dispersion terms become commensurate with the  the group velocity dispersion, such that the integrated dispersion $D_\mathrm{int}(\mu) = \sum_{k>1} \frac{D_k}{k!}\mu^k = \omega_\mr{res}(\mu) - (\omega_\mr{pmp}+ D_1\mu)$ presents multiple zero crossings. %
These zero-crossings are necessarily phased matched with the DKS since $\omega_\mr{res}(\mu_\mr{DW}) = \omega_\mr{dks}(\mu_\mr{DW})$, creating an oscillatory tail bound to the DKS at the azimuthal component $\mu_\mr{DW}$. %
In the OFC, it results in a net gain for the comb lines around $\mu_\mr{DW}$. %
Beyond providing sufficient power in the comb lines of interest to achieve octave-spanning operation, DWs also highlight resonant modes that are nearly phase-matched with the DKS. %
Thus, $D_\mr{int}(\mu_s)$ is minimized at the DW mode, allowing for synchronization to occur, while also letting the reference pump be in the normal dispersion regime where soft-excited bright DKS states cannot be generated. This prevents both multi-DKS states and modulational instability states directly driven by the reference pump.

\begin{figure*}[t]
    \begin{center}
        \includegraphics{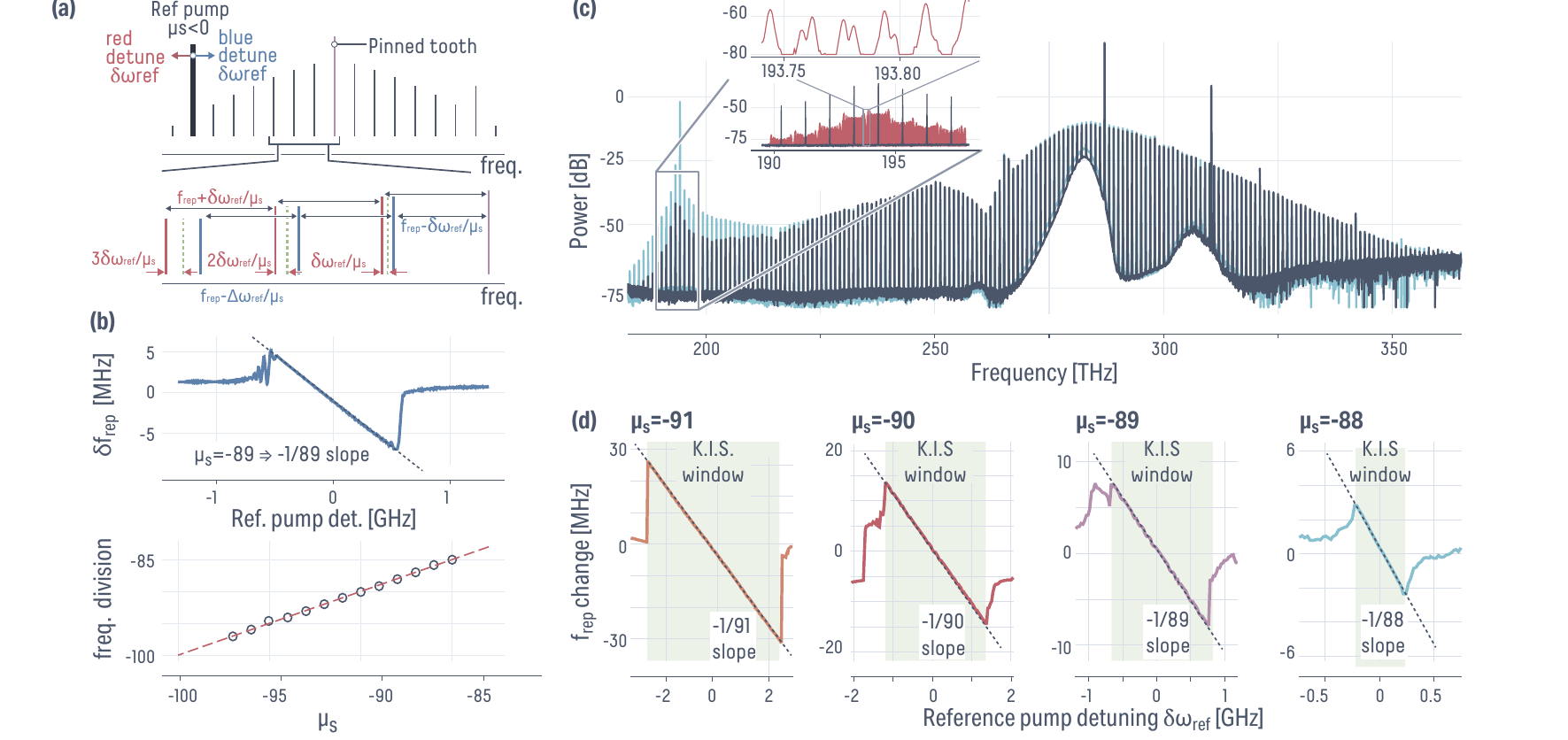}
    \end{center}
    \caption{\label{fig:4} \textbf{Control of the DKS repetition rate from the reference pump alone} -- %
    \textbf{(a)} Concept of repetition rate tuning. In the KIS state, the reference pump is a comb tooth, and thus its tuning stretches or compresses the OFC, resulting in a shift of the repetition rate. \textbf{(b)} LLE simulation of the repetition rate shift with the reference pump detuning. The slope in the synchronization window is the expected frequency division factor $\mu_s = -89$ (top). We verify the change in slope when the reference pump is tuned to different azimuthal mode numbers (bottom). %
    \textbf{(c)} Optical frequency comb under study, similar to \cref{fig:3}, while introducing the electro-optic comb apparatus to frequency divide the DKS OFC repetition rate ($\approx1$~THz) into a measurable frequency. The light blue comb spectrum is in the case of synchronization to the reference pump, while the dark blue is in the case of no reference pump. The inset is a zoom-in on the region near the long DW, with the red trace showing the EO phase modulation of the DKS comb teeth.  %
    \textbf{(d)} Experimental measurement of the repetition rate change with the reference pump detuning. The slope in the synchronization window respects the division factor $\mu_s$ for different probed azimuthal mode numbers, in accordance with simulation predictions. 
    }
\end{figure*}
Our system under study is a resonator with a 23 $\mu$m outer radius that is made of $H=670$~nm thick \ce{Si3N4} with a ring width of $RW=830$~nm embedded in \ce{SiO2}. %
It has an anomalous dispersion when pumped at 286~THz (1048~nm), and this design yields an octave-spanning bandwidth while reaching the rubidium two-photon transition at $\approx 386$~THz\cite{MartinPhys.Rev.Appl.2018,YuPhys.Rev.Applied2019a}. %
We characterize the dispersion of the fundamental transverse electric mode of the cavity using multiple continuously tunable lasers (CTLs) and calibrate each resonance using a wavemeter. %
The obtained experimental $D_\mr{int}(\mu)$ matches with the expected design and finite element method (FEM) simulation, exhibiting a low-frequency DW at $\omega_\mathrm{DW}/2\pi\approx 194$~THz (1544~nm), which corresponds to $\mu_\mr{DW} = -92$ [\cref{fig:3}(a)]. %
Although a high-frequency DW should also exist, the waveguide pulley coupler length $L_\mr{c}=28$~{\textmu}m is chosen to optimize couplings at both the pump and the low-frequency DW for efficient reference pump injection. %
To this end, the high-frequency coupling has been sacrificed, and this DW will not be observed in the experimental frequency comb extracted into the waveguide (though it should exist inside the cavity). %
We actively cool the resonator using a counter-propagative cross-polarized 308~THz (974~nm) laser, allowing for adiabatic detuning of the pump to reach the single DKS state~\cite{ZhangOptica2019a, ZhouLightSciAppl2019,MoillearXiv2023}, which we obtain with an on-chip pump power of about 150~mW. %
To probe the synchronization of the system, we use a CTL which we can park at different resonant modes $\mu$ and adjust its detuning finely to study the impact of $\delta\omega_\mr{ref}$. %
Optically, a large change of behavior is observed with varying $\delta\omega_\mr{ref}$ [\cref{fig:3}(b)]. %
When the reference is precisely on resonance [\cref{fig:3}(b) - on resonance case], the comb lines around $\mu_\mr{DW}$ are maximized and have an offset that is larger than the 4 GHz resolution of the optical spectrum analyzer, This case enables the generation of a 2D comb~\cite{MoillearXiv2023}, and is an unsynchronized state. %
Adjusting $\delta\omega_\mr{ref}$ first reduces these comb lines, decreasing the value of $\Omega = \omega_\mr{ref} - \omega_\mr{\mu,s}$, before the comb lines then increase again, suggestive of a synchronized system. %
Due to the coherent nature of the system and that the DKS comb and the XPM-induced one from the reference pump are merging, the power at each azimuthal component sums up and explains the increases in the comb line power in the overlap region [\cref{fig:3}(b) - sync case]. %
A detuning from the resonance is needed to enter synchronization for two main reasons. %
First, there is the existence of a nonlinear phase shift that requires compensation by $\delta\omega_\mr{ref}$, as predicted by the Adler equation (Eq.~\ref{eq:adler}). %
Secondly, because $D_\mr{int}(\mu)$ is discrete, it will not precisely be zero at the DW mode.

We proceed to record $\Omega$ -- obtained by measuring the comb with a fast photodiode and processed with an electrical spectral analyzer -- with the reference pump detuning $\delta\omega_\mr{ref}$ [\cref{fig:3}(c)], to demonstrate the synchronization dynamics of the system. %
The reference pump is centered at $\mu=-92$ and has $\approx2$~mW of on-chip power. %
Out of synchronization, the OFC that is observed has interleaved comb components, each with a different CEO, induced by $\partial \Phi/\partial t\neq 0$ and as observed in refs.~\cite{MoilleNat.Commun.2021a, MoillearXiv2023, QureshiCommunPhys2022, ZhangNatCommun2020}. %
Experimentally, a beat note $\Omega$ is recorded in the electrical domain between the two components, with a linear dependence on the reference external laser detuning [\cref{fig:3}(c), red dashed line]. %
Higher harmonics $n\times\Omega$ are also expected once $\Omega$ becomes small enough, which is the signature of the recently observed two-dimensional comb~\cite{MoillearXiv2023}. %
Yet, the linear trend becomes hyperbolic once the laser is adjusted sufficiently close to the Kerr-induced synchronization (KIS) window. %
Appleton first observed this behavior in his triode oscillator synchronization experiment~\cite{Appleton1922}, and it is a signature of the onset of synchronization. %
We find that $\Omega$ vanishes, indicating that KIS has occurred, over a range of $\approx1.75$~GHz of reference pump frequency tuning. %
We also find that the Lugiato-Lefever equation (LLE) model under multi-driving fields introduced in ref.~\cite{TaheriEur.Phys.J.D2017} captures this KIS behavior accurately. %
In that work, though the second pump was situated in anomalous dispersion, which greatly limits experimental demonstration in contrast to pumping at the DW mode, the authors theoretically showed that that an external laser different from the main pump may become a comb tooth, allowing for control of the repetition rate. %
Using the \textit{pyLLE} freeware~\cite{MoilleJ.RES.NATL.INST.STAN.2019}, we reproduce the experimental observation of the CEO offset by periodically extracting the DKS every-round trip (as in the experiment) and probing the frequency difference in CEO between the comb components (see Methods). %
We proceed to measure different azimuthal modes $\mu$ for the reference pump [\cref{fig:3}(e)]. %
From the extended Adler equation presented in~\cref{eq:adler}, one can extract the maximum locking window to be $\Delta \omega_\mr{lock} \propto\sqrt{|A_{\mu,s}|^2P_\mr{ref}}$, which is proportional to the reference pump power and also to the $\mu^{th}$ DKS component power, to which the reference pump locks. %
Therefore, increasing the on-chip reference pump power up to about 3~mW increases the reference detuning window for which the system is synchronized. %
Probing modes around the DW enables power variation of the DKS azimuthal component under study. %
As phase matching is obtained at the DW, the neighboring modes will only exhibit lower comb tooth power [\cref{fig:3}(d)]. %
Thus, the KIS window is reduced for the same reference power. %
We note that we have normalized the detuning relative to the center of the KIS window, as under DKS operation, obtaining the exact resonance position is experimentally challenging. %
All the measurements have been performed with a reference pump detuning from red (lower frequency) to blue (higher frequency). %
The same synchronization effect happens with a reversed detuning direction, where a hysteris behavior in the Arnold tongue frequency is observed (see Supplementary Material \cref{sup:Arnold_bistability}).

\begin{figure*}[t]
    \begin{center}
        \includegraphics{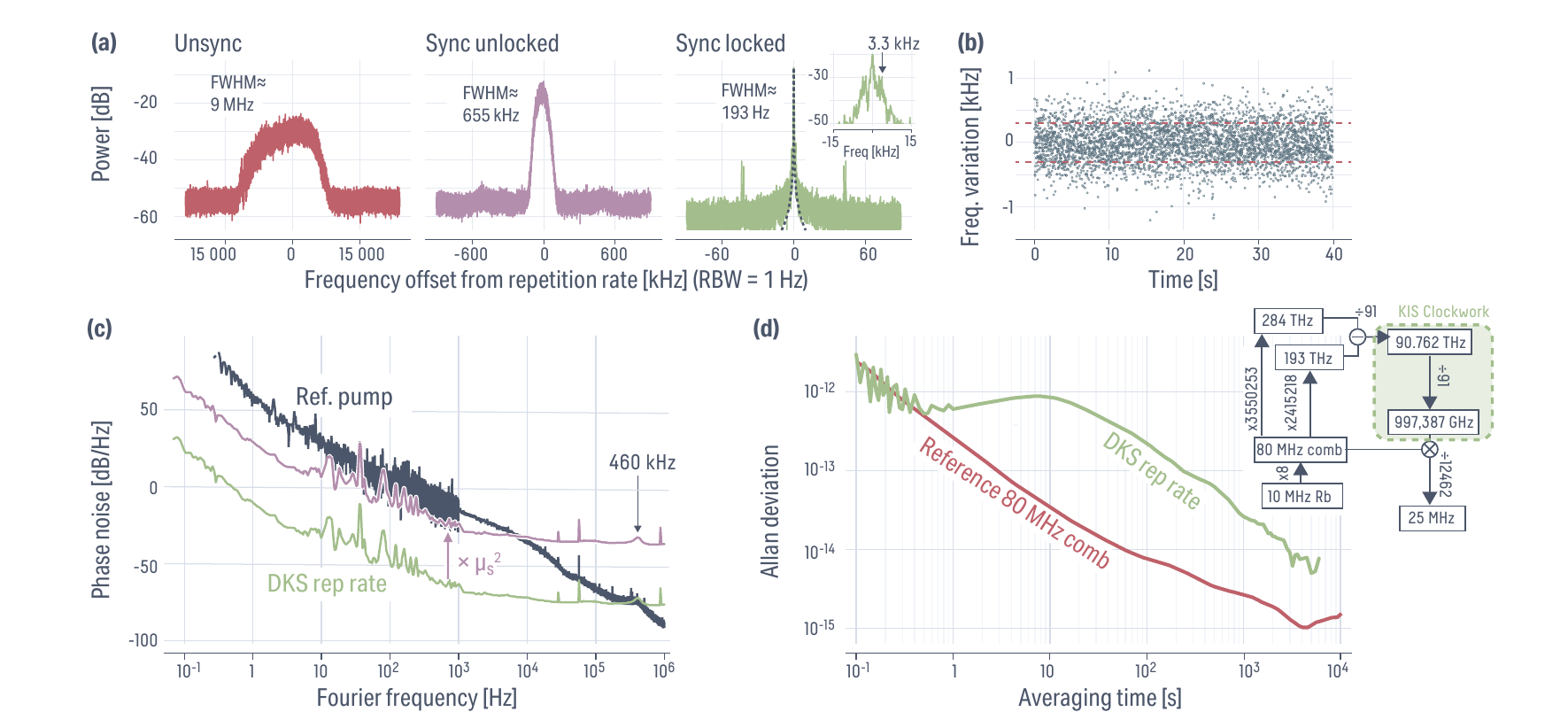}
    \end{center}
    \caption{\label{fig:5} \textbf{Validation of the clockwork under passive synchronization} -- %
    \textbf{(a)} Electrical spectrum analyzer measurement of the DKS repetition rate under different conditions, including without reference pump synchronization (left), synchronization but with the main and reference unlocked (middle), and synchronized with the main and reference pumps locked to a fiber frequency comb. The frequency axes have been referenced to the actual DKS repetition rate frequency. The power is normalized to 1~mW. %
    \textbf{(b)} Frequency counter trace {(gate time of 10~ms)} of the repetition rate in a synchronized state with lasers locked to the fiber comb, highlighting a standard deviation of about 200 Hz (red dashed lines), hence a repetition rate of 997,387,327,768~Hz~$\pm~200$~Hz. 
    \textbf{(c)} Phase noise of the reference pump (blue-gray) and repetition rate for the synchronized locked state to the fiber comb (green). The phase noise of the reference pump is divided onto the repetition rate. The purple trace shows that the repetition rate noise, multiplied by the frequency division factor $\mu_s^2$, is close to that of the reference pump. Power is referenced to that of the carrier, i.e., dBc/Hz. %
    \textbf{(d)} Long term stability of the repetition rate in the synchronized locked state (green) compared to the fiber comb (red) to which the main and reference pump are locked. The $1/\tau$ trend at long averaging time demonstrates the coherent phase locking and efficiency of the passive synchronization as a clockwork. The plateau at 1~s to 10~s is thought to come from environmental fluctuations. The schematic indicates the links between 10~MHz Rb frequency standard, 80~MHz fiber frequency comb, and the measured DKS repetition rate under synchronization. %
    }
\end{figure*}

Since $\partial \Phi /\partial t = 0$ in the KIS state, any fluctuations in the reference frequency are divided onto the repetition rate such that $\partial(\delta \omega_\mr{ref})/\partial \omega = \mu_s \partial\omega_\mr{rep}/ \partial \omega$,  assuming that the main pump is fixed (see Supplementary Material Section ~\ref{sup:adler}). %
As a result, the azimuthal mode difference $\mu_s$ between the primary pump and the reference pump defines the frequency division factor in this dual-pinned comb. %
Therefore, by adjusting the detuning of the reference pump while synchronizing with the DKS, one could adjust its repetition rate without interfering with the main pump parameters [\cref{fig:4}(a)]. %
Although the extended Adler equation itself provides a prediction of such frequency division and disciplining of the repetition rate with the reference pump detuning, we verify this behavior directly using the LLE model [\cref{fig:4}(b)]. %
We tune the reference pump around the azimuthal mode $\mu_s = -89$ and extract the repetition rate (see Methods for the simulation protocol). %
When the KIS state is reached, the closest comb tooth to the reference pump is pulled to exactly match the reference pump frequency, creating a sharp modification of $\omega_\mr{rep}$. %
The DKS repetition rate then follows a linear trend with the reference detuning up to the edge of the synchronization window. %
The repetition rate slope with the reference pump detuning follows the expected value for frequency division, which is the number of azimuthal modes between the main ($\mu_p =0$) and reference pump modes ($\mu_s =-89$). %
We verify the slope for different azimuthal modes and find that the slope always equals the frequency division rate $\mu_\mr{s}$ expected from the extended Adler equation. \\
\indent In order to experimentally demonstrate such an effect, we must accurately measure the repetition rate of the OFC, which is generally not directly achievable with an octave-spanning microcomb. %
For an OFC to span over an octave in an on-chip format, the repetition rate must be high, here about 997~GHz, to minimize the number of comb teeth needed to span the octave and reduce the pump power needed. %
We use an electro-optic comb apparatus~\cite{StonePhys.Rev.Lett.2020} made of three cascaded electro-optic (EO) phase modulators driven at $f_\mr{EO}=17.47$~GHz to measure the DKS repetition rate by modulating two of its adjacent OFC comb teeth. %
The phase modulation of the DKS OFC teeth forms an EO comb spanning over 1~THz with $N_\mr{EO}=57$ EO teeth, which, where they overlap, allows us to record their beat note and infer the DKS repetition rate $f_\mr{rep} = N_\mr{EO}f_\mr{EO} \pm f_\mr{beat}$ [\cref{fig:4}(c)]. %
Using this repetition rate measurement method, we demonstrate that the slope of the linear trend of the repetition rate shift with the reference frequency matches the mode difference between the DKS pump and the reference pump, and is in accordance with the frequency division principle for several azimuthal modes [\cref{fig:4}(d)]. %
The KIS windows are different for each azimuthal mode, as the comb tooth power in the DKS is not constant with the probed $\mu$. %
We note that adjusting the reference pump frequency in this dual pinned system not only affects the repetition rate, but also tunes the CEO frequency, and as a result may make it possible to adjust the CEO of the frequency comb so that it falls within a photodetector's detection bandwidth. In addition, it is worth discussing how our synchronization of the DKS to the reference pump is related to recent work on injection locking of lasers for turn-key DKS generation~\cite{ShenNature2020a,VoloshinNatCommun2021}. %
In such injection locking, the back-scattered signal from the DKS resonator is sent back into the (isolator-free) laser system, selecting and locking the laser frequency to one appropriate for DKS generation. %
In the case we present here, the reverse effect takes place. %
All lasers and amplifiers in use are paired with isolators, and hence the synchronization effect only takes place within the microresonator itself (there is no impact on the lasers), hence the Kerr-induced term. %
Instead of the laser frequency adjusting for locking, the reference pump is kept at fixed frequency while the DKS adjusts the frequency of its azimuthal components to fulfill the locking condition. %
The experimental demonstration of the repetition rate tuning highlights this effect.

\vspace{1ex}
Finally, we experimentally demonstrate the viability of the passive Kerr-induced synchronization approach for clockwork applications. %
Although demonstrating frequency division, the main prerequisite for using a microcomb in a clockwork, has been accomplished by measuring the repetition rate tuning with the reference pump frequency [\cref{fig:4}], demonstration of short-term and long-term stability and its division against the reference must also be verified. %
In particular, we show that the phase noise and the long-term stability of the reference pump are divided onto the repetition rate. %
As noted earlier, the inability to extract the short wavelength DW of the DKS OFC prevents us from stabilizing its CEO directly, e.g., via the $f-2f$ method. %
Instead, we use a Toptica Photonics 80~MHz fiber comb that achieves CEO-free operation through difference frequency generation~\cite{LiehlPhys.Rev.A2020, NISTdisclaimer}, referenced to a 10~MHz Rubidium frequency standard, so that we can pin the main and reference pumps in a phase-coherent fashion (i.e., their noise is correlated). %
Rather than the full frequency division if the DKS CEO is locked, a frequency division factor $\mu_\mr{s}$ between the two pinned comb lines is achieved. %
First, we demonstrate that the repetition rate is indeed much more stable in the synchronized case [\cref{fig:5}(a)]. %
Without any synchronization, the repetition rate beat note measured with the EO comb apparatus is relatively unstable, with a full-width at half-maximum (FWHM) of several MHz. %
When the DKS is synchronized to the reference laser but both lasers are unlocked, the repetition rate is more stable, but still exhibits a large FWHM of about 655~kHz, due to the uncorrelated noise of both the main and reference pumps. %
However, when both pumps are additionally locked to the fiber comb and become phase coherent, the repetition rate beat note drastically reduces to a FWHM of about 193~Hz, with a perfect Lorentzian shape, albeit with the appearance of shoulders at 3.3~kHz. %
They can be explained as the frequency divided servo loop bandwidth, which is expected at 460~kHz [\cref{fig:5}(c)], and is consistent with the frequency division given by the number of fiber comb teeth within two adjacent DKS OFC teeth of $N_\mr{fiber}=12462$, the $N_\mr{EO} = 57$ EOcomb teeth between the two DKS OFC teeth, and the mode spacing $\mu_{s}=91$ between the two pumps that perform the principle frequency division (\textit{i.e.}, 460~kHz $\times~\mu_s/(N_\mr{fiber}\times N_\mr{EO}) = 3.3$~kHz). %
We note that all the spectral frequencies in Fig.~\ref{fig:5}(a) have been referenced to the actual DKS repetition rate of about 997,387,327,768~Hz~$\pm~200$~Hz. The error in the repetition rate is estimated by counting its frequency in time and assuming one standard deviation from its average, with a gate time of 10~ms~[\cref{fig:5}(b)]. \\
\indent We measure the phase noise of the repetition rate and the reference pump to demonstrate the short-term stability division. %
The main pump phase noise is similar to the reference one, given that they are both locked to the same fiber comb system and are the same type of CTL laser. %
We measure the phase noise of the reference pump using a 10~MHz Mach-Zehnder interferometer (see supplementary material for protocol~\cref{sup:mzi}) and the repetition rate noise using a phase noise analyzer (PNA) referenced with the 10~MHz Rubidium standard. %
The repetition rate exhibits a much lower phase noise than the reference pump [Fig.~\ref{fig:5}(c)]. %
The two phase noise measurements almost match when accounting for the frequency division factor $\mu_s^2$. %
A slight discrepancy remains, and the better performance of the repetition rate against the divided reference is likely due to the mutual coherence between the main and reference pumps; \textit{i.e}, because the two pumps are locked to the same reference comb, their absolute frequency fluctuations are correlated, and this common noise is rejected by the repetition frequency.\\
\indent To fully demonstrate the potential of the passive synchronization approach for clockwork application, we measure the long-term stability of the system by measuring the repetition rate Allan deviation [\cref{fig:5}(d)]. %
The $1/\tau$ trend at long averaging time indicates that the only contributions are from white phase noise and demonstrates the phase coherence with the 80~MHz fiber comb. %
We show a residual noise of 1 part in $1.25\times10^{12}$ at 1 second, similar to the fiber comb stability at 1 second and demonstrating the stability division, and 1 part in $10^{14}$ at 50 minutes. %
The plateau appearing between 1 second and 10 seconds is likely due to environmental fluctuations, which we have not attempted to reduce and is consistent with previous observations of active locking of a microcomb clockwork~\cite{DrakePhys.Rev.X2019}. %
We note that the long-term stability metric is likely limited by a number of factors, including the complexity of our frequency division experiment (which includes both multiplication and division steps [\cref{fig:5}(d)]) and the stability of the fiber comb source and its Rb frequency standard. In the future, we anticipate that much better stability may be obtained using a two-photon stabilized reference laser and a CEO-locked main pump, whereby full frequency division from the optical to microwave can be realized without intermediate multiplication steps.

In conclusion, we have demonstrated a new approach for a microcomb-based optical frequency division clockwork. Rather than active stabilization of the microcomb to a reference laser, our approach is based on passive, all-optical Kerr-induced synchronization (KIS) that bypasses electronics that add to the complexity and power budget of the optical frequency divider. In analogous fashion to mechanical clockworks, passive synchronization is enabled by direct coupling to the reference. In particular, the same optical nonlinearity that makes DKS generation possible is used to achieve strong synchronization between the DKS and the reference laser, bypassing the need for active stabilization. By experimentally and theoretically investigating the nonlinear dynamics of the system and demonstrating KIS with as little as 2~mW of on-chip reference {pump} laser power, compatible with on-chip integrated lasers~\cite{XiangScience2021}, we demonstrate passive synchronization of a microcromb clockwork for the first time. Indeed, in contrast to existing self-injection locking approaches where the main pump laser frequency follows the DKS existence condition, here it is the DKS which locks its frequency components to a separate reference {pump}, providing a direct external control to tune the microcomb parameters. We show that the repetition rate of the DKS can be controlled by adjusting the reference {pump} frequency, with a tuning set by the frequency division factor. Finally, by demonstrating short- and long-term stability division compatible with an optical clock architecture, we show the viability of our clockwork technique. This synchronization technique can be straightforwardly extended to a fully self-referenced microresonator frequency comb with improved coupling engineering to efficiently extract the short wavelength DW. The reference pump being a comb tooth could provide substantially larger power than the originally available comb tooth at the DW for second harmonic generation, to enable the nonlinear interferometry in a more efficient fashion, and allowing for optimized optical frequency division. Our work is a key milestone toward compact optical clocks for fieldable applications in position, navigation, and timing, both by decreasing the power budget and greatly simplifying the {architecture}. It also paves the way for future physics investigations that can probe novel DKS states thanks to the nonlinear connection between CW light and intra-cavity pulses.

                               
\bibliographystyle{apsrev4-2}
%

\section*{Methods}

\textbf{LLE simulations -- }
The extended Lugiato Lefever equation model including multiple driving fields can be written as \cite{TaheriEur.Phys.J.D2017}: 

\begin{align}
    \label{eq:mLLE}
\frac{\partial a(\theta, t)}{\partial t } &= -\frac{\kappa}{2} a + i\sum_\mu A(\mu, t) D_\mr{int}(\mu) \mr{e}^{i\theta\mu} \\
& + \Delta D_1\frac{\partial a}{\partial \theta}  + i\gamma L |a|^2a \nonumber\\
& + i\sum_j \sqrt{\kappa_\mr{ext} P_j} \mr{e}^{i\mu_j \theta + i(D_\mr{int}(\mu_j)+ \delta\omega_j)t} \nonumber
\end{align}

\noindent with $a(\theta, t) = \sum_\mu A(\mu, t)\mr{e}^{i\theta\mu}$, $\kappa$ and $\kappa_\mr{ext}$ the total and external losses, $P_j$ the power of each driving field $j$, $D_\mr{int}(\mu_j)$ the integrated dispersion at the azimuthal mode $j$, $\gamma = \omega \frac{n_2}{A_\mr{eff}}$ the effective nonlinearity with $n_2$ the material nonlinearity and $A_\mr{eff}$ the effective mode area, $L=2\pi R$ the resonator round trip length, and  $\Delta D_1$ the mismatch between the linear free spectral range at the pump (\textit{i.e} from which $D_\mr{int}$ is defined) and the soliton repetition rate. In our system, we are in the presence of two driving fields, the main pump at $\mu = 0$ and the reference at $\mu_s$. To obtain the optical frequency comb that is observed in experiment the following protocol is observed, similar to the one presented in ref \cite{MoillearXiv2023}: 
\begin{enumerate}[topsep=0pt,itemsep=-1ex,partopsep=1ex,parsep=1ex]
    \item For each reference detuning $\delta\omega_\mr{ref}$, find $\Delta D_1$ such that the soliton position in $\theta$ is invariant with $t$
    \item Solve the LLE and sample the soliton at every round trip time to effectively emulate the periodic extraction by the bus waveguide 
    \item Reconstruct the pulse train by concatenating every sampled soliton to obtain $a_\mr{wg}(t)$
    \item Fourier transform the pulse train to reconstruct the OFC and obtain $A_\mr{wg}(\omega)$
    \item Recast the frequency comb into a two dimensional system $A_\mr{wg}(\mu, \Omega)$ by resizing $A_\mr{wg}(\omega)$ by the repetition rate sampling size, thus any power component within $[-\omega_\mr{rep}/2; +\omega_\mr{rep}/2]$ around a given comb tooth $\mu$ of the DKS 
    \item Integrate $A_\mr{wg}(\mu, \Omega)$ along $\mu$ to obtain $S(\Omega)$, which is equivalent to the experimental measurement of the beat note arising from the CEO frequency shift. 
    \item Repeat for every $\delta\omega_\mr{ref}$ to create the plot presented in \cref{fig:3}(c). 
\end{enumerate}

In order to retrieve the dynamics presented in \cref{fig:4}, there is no need to obtain $S(\Omega)$ but rather it is about finding $\Delta D_1$. If one assumes a very slow variation of $\delta\omega_\mr{ref}$, we can make the approximation that $\delta\omega_\mr{ref}$ is constant in between $t$ and $t+\delta t$. One can then extract the drift of the soliton within this time window and obtain $\Delta D_1$. 

\vspace{1ex}

\textbf{Experimental Protocol --}
For a complete description of the experimental setup, please refer to the supplementary material~\cref{sup:setup}. To measure the CEO offset highlighting the synchronization in \cref{fig:2}, a 6~GHz photodiode measures the comb, with the main pump and the cooler filtered out. For measuring the repetition rate in \cref{fig:4}, we send two adjacent OFC teeth through the EO comb apparatus. We then use a narrow filter to only select the two EO comb teeth that participate in the frequency downconversion repetition rate beat note, which we record with a 50~MHz avalanche photodiode. For the noise measurements presented in \cref{fig:5}, the same type of filtering is used with the 80~MHz fiber comb beating against a single DKS OFC tooth. 
\\

\noindent \textbf{\large Data availability} \\
The data that supports the plots within this paper and other findings of this study are available from the corresponding authors upon reasonable request. The simulation code is available from the authors through the pyLLE package available online~\cite{MoilleJ.RES.NATL.INST.STAN.2019}.\\

\noindent \textbf{\large Author Contributions} \\
G.M. led the project. J.~S. helped with the metrology analysis. M.~C. developed the electro-optic comb. C.~M. contributed in the understanding of the physical phenomenon. K.~S. helped with guiding the project and in data analysis. G.~M. and K.~S. wrote the manuscript, with input from all authors. All the authors contributed and discussed the content of this manuscript. \\

\noindent \textbf{\large Acknowledgments} \\
The photonic chips were fabricated by Ligentec SA. The authors acknowledge partial funding support from the AFRL Space Vehicles Directorate, the DARPA APHI program, and the NIST-on-a-chip program. We thank Pradyoth Shandilya, Marcelo Davan\c{c}o, and Vladimir Aksyuk for insightful feedback. We thank Michael Highman and Matt Cich from Toptica Photonics for loan of the fiber comb.\\

\noindent \textbf{\large Competing Interests} \\
The authors declare no competing interests.

\clearpage
\beginsupplement

\section*{Extended data}
\section{Adler equation for the Multi-Color DKS (MDKS)}
\label{sup:adler}
We are seeking to find the equation of motion of the system, in particular the equation describing the evolution in time of the phase difference between the reference pump laser $\varphi_\mathrm{ref}$ at azimuthal mode $\mu_s$ (defined relative to the main pump at $\mu=0$) and the corresponding DKS azimuthal component $A_{\mu_s}$ with the phase $\varphi_\mathrm{dks}(\mu_s)$. We know that the carrier envelope offset frequency arises from the phase velocity, such that $\omega_\mathrm{ceo} = \partial \varphi /\partial t$. We are assuming that the pulse that is generated by the reference laser is locked to the pulse that is generated by the pump laser to form a single MDKS that propagates with a single group velocity. However, each component of the MDKS will in general have its own phase velocity, hence its own CEO frequency $\omega_\mathrm{ceo}^\mathrm{(ref)}$ and $\omega_\mathrm{ceo}^\mathrm{(dks)}$ for the reference and main component of the MDKS respectively. This phase velocity difference is given by:
\begin{align}
    \label{eq:phase}
    \frac{\partial \Phi}{\partial t} &= \frac{\partial \varphi_\mr{ref}}{\partial t}  - \frac{\partial \varphi_\mr{dks}(\mu_s)}{\partial t} \nonumber\\
    & = \omega_\mr{ceo}^\mr{( ref)}  - \omega_\mr{ceo}^\mr{( dks)}\nonumber\\
    & = \omega_\mr{ref}  - \omega_\mr{dks}(\mu_s)\\
    & = \omega_\mr{cav}(\mu_s) + \delta\omega_\mr{ref} - (\omega_\mr{0} + \mu_s \omega_\mr{rep}) \nonumber
\end{align}

\noindent with $\varphi_\mr{ref}$ and $\varphi_\mr{dks}(\mu_s)$ the relative phase of the reference pump and its closest DKS azimuthal component respectively, $\omega_\mathrm{ceo}^{(X)}$ being the CEO frequency relative to the DKS ($X=\mathrm{dks}$) or the reference pump ($X=\mr{ref}$), $\omega_\mr{ref}$ is the frequency of the reference pump which is offset from the cavity resonance $\omega_\mr{cav}(\mu_s)$ by a detuning $\Delta \omega_\mr{ref}$, and $\omega_\mr{dks} (\mu_s)= \omega_\mr{0} + \mu_s  \omega_\mr{rep}$ is the frequency of the closest comb tooth from the reference pump frequency with $\omega_\mathrm{rep}$ being the repetition rate of the DKS and $\omega_0$ being the main pump frequency.  $\frac{\partial \Phi}{\partial t}$ represents the repetition rate in the second dimension of the 2D comb reported in ref.~\cite{MoillearXiv2023}. 

Additional differentiation of~\cref{eq:phase} provides a direct link between the phase mismatch $\Phi$ and $\omega_\mr{rep}$:
\begin{align}
    \label{eq:d2phi}
    \frac{\partial^2 \Phi}{\partial t^2} &= - \mu_s\frac{\partial \omega_\mr{rep}}{\partial t} 
\end{align}

We now suppose that in the resonator angular coordinate $\theta$ there is a short-duration pulse whose central angle $\bar{\theta}$  is well-defined and rotates at a constant angular velocity with a fixed shape. Therefore, we have:

\begin{equation}
 E_\mr{dks} \bar{\theta} = \int_{-\pi}^\pi \theta |a(\theta, t)|^2 \frac{\mr{d}\theta}{2\pi} 
\end{equation}

\noindent where $a(\theta, t)$ satisfies the modified LLE equation presented in \cref{eq:mLLE}. The quantity $E_\mr{dks}$ is given by $\int\frac{\mr{d}\theta}{2\pi} |a(\theta, t)|^2$ and represent the DKS energy, so that $\bar{\theta}$ is the pulse averaged central angle of the pulse. We then have 

\begin{equation}
    \label{supeq:dUdt}
    E_\mr{dks} \frac{\mr{d}\bar{\theta}}{\mr{d}t} = \frac{\mr{d}}{\mr{dt}}{\int_\pi^\pi  \theta |a(\theta, t)|^2} \frac{\mr{d}\theta}{2\pi} = \int_{-\pi}^\pi  \theta \Big[a\frac{\partial a^*}{\partial t} + a^*\frac{\partial a}{\partial t} \Big]  \frac{\mr{d}\theta}{2\pi}
\end{equation}
We must first assume that the loss and gain terms make a negligible contribution to the momentum. This assumption implies:

\begin{equation}
    0 \approx \int_{-\pi}^\pi  \theta |a(\theta, t)|^2 \frac{\mr{d}\theta}{2\pi}= \int_{-\pi}^\pi  \theta \Big[Fa^* + F^* a - \kappa |a|^2\Big] \frac{\mr{d}\theta}{2\pi}
\end{equation}
with $F$ being the main driving force in \cref{eq:mLLE} and we assume the reference pump is negligible before the main pump power, which is the case in our experiment. The assumption that the loss and gain terms are negligible is reasonable since the magnitude of these terms is small. There is, however, no expectation that it is exactly zero. Otherwise, the Kerr effect terms cancel out, so we can thus focus on the linear contribution, which is the dominant contribution. It is useful now to work in the mode number domain. We write:

\begin{equation}
    A(\mu, t) = \int_\pi^\pi \frac{\mr{d}\theta}{2\pi} a(\theta, t)\mr{e}^{-i\mu\theta}, \quad a(\theta, t) = \sum_\mu A(\mu, t)\mr{e}^{i\mu t}
\end{equation}

We then have under the same approximations:
\begin{equation}
    \label{supeq:dAmu}
    \frac{\mr{d}A(\mu)}{\mr{d}t} \equiv  i\omega_\mr{cav}(\mu) A(\mu)
\end{equation}

\noindent with $\omega_\mr{cav}(\mu)$ the azimuthal component optical frequencies on resonance. Using \cref{supeq:dAmu} and injecting it in \cref{supeq:dUdt}, accounting only for the linear terms, integrating by parts, and noting that only the term $\mu = 0$ survive the integration, we obtain:
\begin{equation}
    E_\mr{dks} \frac{\mr{d}\bar{\theta}}{\mr{d}t} = -\sum_\mu \omega'_\mr{cav}(\mu) A(\mu) A^*(\mu) 
\end{equation}
with $\omega'_\mr{cav}(\mu) = \frac{\mr{d}\omega_\mr{cav}(\mu)}{\mr{d}\mu}$. It should be noted that we are assuming that $A_\mu$ and $\omega_\mr{cav}(\mu)$ are both continuous functions of $\mu$ with continuous derivatives, and thus assuming a resonator without mode coupling to different propagation directions (\textit{i.e.} backscattering) or mode families (\textit{i.e} avoided mode crossing). 

The dispersion can be accounted in different fashions, either through a Taylor expansion assuming mode-independent coefficients or assuming a mode-dependent free spectral range such that $\omega_\mu' = \frac{\mr{d}\omega_\mu}{\mr{d}\mu} = D_1(\mu)$. Noting that in the cavity referential the temporal variation of the the pulse averaged central angle $\frac{\partial\bar\theta}{\partial t}$ translates directly into the repetition rate $\omega_\mr{rep}$, we can write : 

\begin{equation}
    \omega_\mr{rep} =  - \frac{1}{E_\mr{DKS}}\sum_\mu D_1 (\mu) A(\mu) A^*(\mu)
\end{equation}

Taking the time derivative of this equation, we obtain:

\begin{equation}
    \label{eq:omega_rep_dot}
    \frac{\partial\omega_\mr{rep}}{\partial t} = -\frac{1}{E_\mr{DKS}}\sum_\mu D_1 (\mu) \left( \frac{\partial A(\mu)}{\partial t} A^*(\mu) + A(\mu) \frac{\partial A^*(\mu)}{\partial t} \right)
\end{equation}

We can write the coupled mode theory equation for each azimuthal mode component $A(\mu)$ accounting for the Kerr nonlinearity as:

\begin{equation}
    \label{eq:CMT}
    \frac{\partial A(\mu)}{\partial t} = \left(-\frac{\kappa}{2}  + i \omega_\mr{cav}(\mu) \right)A(\mu) + i \gamma L\sum_{\alpha, \beta}A(\alpha) A^*(\beta) A(\alpha-\beta + \mu)  - i\delta(\mu_0) \sqrt{\kappa_\mr{ext} P_\mr{main}}\mr{e}^{i \omega_\mr{0} t} - i\delta({\mu_s}) \sqrt{\kappa_\mr{ext}P_\mr{ref}}\mr{e}^{i \omega_\mr{ref} t}
\end{equation}

\noindent Here, $\omega_\mr{cav}(\mu)$ are the resonance frequencies of the modes $\mu$ normalized to the pump mode, $\gamma = \omega n_2/ A_\mr{eff}$ is the effective nonlinear coefficient, with $n_2 = 2.4\times 10^{-15}$~cm\textsuperscript{2}$\cdot$W\textsuperscript{-1}, which we assume constant for simplification (yet given the comb bandwidth it should not be), $L$ is the circumference of the resonator, $\kappa_\mr{ext}$ is the coupling rate and $\kappa$ the total loss rate, $P_\mr{main}$ and $P_\mr{ref}$ are the main and reference pump laser power, $\mu_s$ is the azimuthal mode at which the reference laser is situated, and $\delta(x)$ is the Kronecker delta function such that $\delta(\mu = x) = 1$ and $\delta(\mu \neq x) = 0$. We note that $\mu_0$ is the main pumped mode, which is null here per normalization, yet we explicitly label it here to avoid confusion with a simple zero.

Plugging~\cref{eq:CMT} into \cref{eq:omega_rep_dot}, several things can be simplified. First, we can note that all linear imaginary components of \cref{eq:CMT} cancel each other out for each $\mu$, and only the terms where the pumps are presents need to be kept. In addition, we can note that the relative phase between the main pump and its corresponding phase is fixed. Thus  we obtain the following equation for the temporal derivative of the repetition rate:

\begin{align}
    -\frac{\partial\omega_\mr{rep}}{\partial t} = \kappa D_{1}(\mu_{s}) K_{s}  \sin{\Big[\big(\varphi_\mr{ref} - \varphi_\mr{dks}(\mu_s)\big) t \Big]} + \kappa D_{1}(\mu_0) K_{0} + \kappa D_{1}(\mu_0) K_{NL} \ - \kappa\omega_\mr{rep} 
\end{align}

\noindent Here, we note that the $\sin$ function appears because of $-i(c-c^*) = 2\mathcal{I}(c)$. The following parameters are defined: 

\begin{minipage}[l]{0.3\linewidth}
    \begin{align}
        K_{s}&=\frac{2 \left|{A(\mu_{s})}\right|}{E_\mr{DKS}} \sqrt{\frac{P_\mr{ref} \kappa_\mr{ext}}{\kappa^{2}}}\nonumber\\
        K_{0}&=\frac{2 \left|{A(\mu_{0})}\right|}{E_\mr{DKS}} \sqrt{\frac{P_\mr{main} \kappa_\mr{ext}}{\kappa^{2}}}\nonumber
    \end{align}
\end{minipage}
\begin{minipage}[c]{0.7\linewidth}
    \begin{align}
        K_{NL}&= \frac{\gamma L}{\kappa D_1(\mu_0)E_\mr{DKS}}\sum_\mu D_1(\mu)\left(A^*(\mu) \sum_{\alpha, \beta}A(\alpha) A^*(\beta) A(\alpha-\beta + \mu) - c.c \right) \nonumber
    \end{align}
\end{minipage}

Using \cref{eq:d2phi} and \cref{eq:omega_rep_dot}, we arrive at: 
\begin{align}
    \frac{1}{\kappa} \frac{\partial^2\Phi}{\partial{t}^2}=- \mu_{s}D_{1}(\mu_{s}) K_{s}  \sin{\left(\Phi \right)} - \mu_{s}  D_{1}({\mu}_{0}) K_{NL} + \mu_{s} \omega_\mathrm{rep} 
\end{align}

Using \cref{eq:phase}, we can replace $\omega_\mr{rep}$ to obtain the extended Adler equation which describes the phase locking of the soliton to the reference laser:

\begin{align}
    -\frac{1}{\kappa} \frac{\partial^2\Phi}{\partial{t}^2} +   \frac{\partial\Phi}{\partial{t}}=- \mu_{s}D_{1}(\mu_{s}) K_{s} \sin{\left(\Phi \right)} - \mu_{s} D_{1}({\mu}_{0}) K_{NL}  - \mu_{s}D_{1}({\mu}_{0}) K_{0}  + \delta\omega_\mathrm{ref} - \omega_{0} + \omega_\mathrm{cav}(\mu_s)
\end{align}

We can recall: 

\begin{align} 
    \omega_\mathrm{cav}(\mu_s) - \omega_0  &= D_\mathrm{int}(\mu_s) + \mu_s D_{1}({\mu}_{0})\nonumber 
\end{align}

Leading to 
\begin{align}
    -\frac{1}{\kappa} \frac{\partial^2\Phi}{\partial{t}^2} - \frac{\partial\Phi}{\partial{t}} + = D_\mathrm{int}(\mu_s) + \Delta - \mathcal{T}\sin{\left(\Phi \right)}
\end{align}

\noindent This equation is similar to the equation of motion of a damped pendulum under constant torque driving~\cite{CoulletAmericanJournalofPhysics2005}, with the effective detuning $\Delta = \delta\omega_\mr{ref} + \mu_sD_1(\mu_0)(1-K_\mathrm{NL} - K_0)$ along with $ D_\mathrm{int}(\mu_s)$ associated to the damping of the effective pendulum system, and $\mathcal{T} = D_{1}(\mu_{s}) K_{s} \mu_{s}$ the effective ``natural'' torque of the system (\textit{i.e.} the mass, cord length and gravity acceleration product). For the synchronization to occur, the right hand-side of the equation needs to cancel out. This leads to a condition on the integrated dispersion at the reference pumped mode, which needs to be minimized or needs to be compensated for by the reference pump detuning. As $\delta\omega_\mr{ref}$ is limited, this imposes a hard limit on which mode can be used. Hence, sending the reference laser at the modes where the DW(s) occur allows one to minimize $D_\mr{int}(\mu_s)$ and reach a synchronization regime.

From the above, we can extract the locking bandwidth for which the reference laser can be tuned while maintaining synchronization, around a center frequency yet to be determined, as follows: 
\begin{align}
    \Delta\omega_\mr{lock} = 2 \mathcal{T}  = 4 D_{1}(\mu_{s}) \mu_{s} \frac{\left|{A(\mu_{s})}\right|}{E_\mr{DKS}} \sqrt{\frac{P_\mr{ref} \kappa_\mr{ext}}{\kappa^{2}}}
\end{align}

To give some more insight about the typical values associated with K.I.S. of the DKS to the reference pump laser in our system, we extract the following parameters from the LLE and experiment: 

\begin{table}[H]
    \begin{center}
        \begin{tabularx}{0.25\linewidth}{c c}
            \toprule
            Parameter  & Values\\
            \midrule
            $|A(\mu_0)|$  & $3\times 10^{2}$ J\textsuperscript{1/2} \\
            $|A(\mu_s)|$  & $1\times10^{-2}$ J\textsuperscript{1/2} \\
            $E_\mr{DKS}$ & $1\times10^{-3}$J  \\
            $\kappa_\mr{ext}$  &  2$\pi~\times$~80~MHz\\ 
            $\kappa$  & 2$\pi~\times$~150~MHz\\ 
            $\mu_s$  & 92\\ 
            $P_\mr{main}$ & 150~mW \\
            $P_\mr{ref}$ & 1~mW \\
            $K_{\mu s}$  & $4\times 10^{-6}$\\
            $K_{0}$  & $1.4 \times 10^{-5}$\\ 
            $D_1(\mu_0)$  & 2$\pi~\times$~1~THz\\
            $\mu_sD_1(\mu_0)K_{0}$  & 2$\pi~\times$~1.26~GHz\\
            $\mu_sD_1(\mu_0)K_{NL}$  & 2$\pi~\times$~100~MHz\\
            $\mathcal{T}$  & 2$\pi~\times$~365~MHz\\
            $\Delta\omega_\mr{lock}$ & 2$\pi~\times$~730~MHz\\
            \bottomrule
        \end{tabularx}
    \end{center}
\end{table}
\vspace{-1ex}

Finally, we can determine how a small change in the reference pump detuning $\delta\omega_\mathrm{ref}$, which we call $\mathrm{d}\delta\omega_\mathrm{ref}$, is related to a small change in $\omega_\mathrm{rep}$, which we call $\mathrm{d}\omega_\mathrm{rep}$, simply by differentiating \cref{eq:phase}, noting that in the synchronization regime $\partial\Phi/\partial t = 0$, and assuming the frequency of resonance and the pump are fixed (as in the experiment). This yields:

\begin{equation}
    \mathrm{d}\delta\omega_\mr{ref} = \mu_s \mathrm{d}\omega_\mr{rep}
\end{equation}

\noindent Hence, any variation of the reference pump detuning is frequency divided onto the repetition rate with a factor defined by the reference pump mode number $\mu_s$ (assuming the main pump remains fixed), similar to the experimental demonstrations in \cref{fig:4}

\section{Complete experimental setup}
\label{sup:setup}
\begin{figure}[H]
    \begin{center}
        \includegraphics{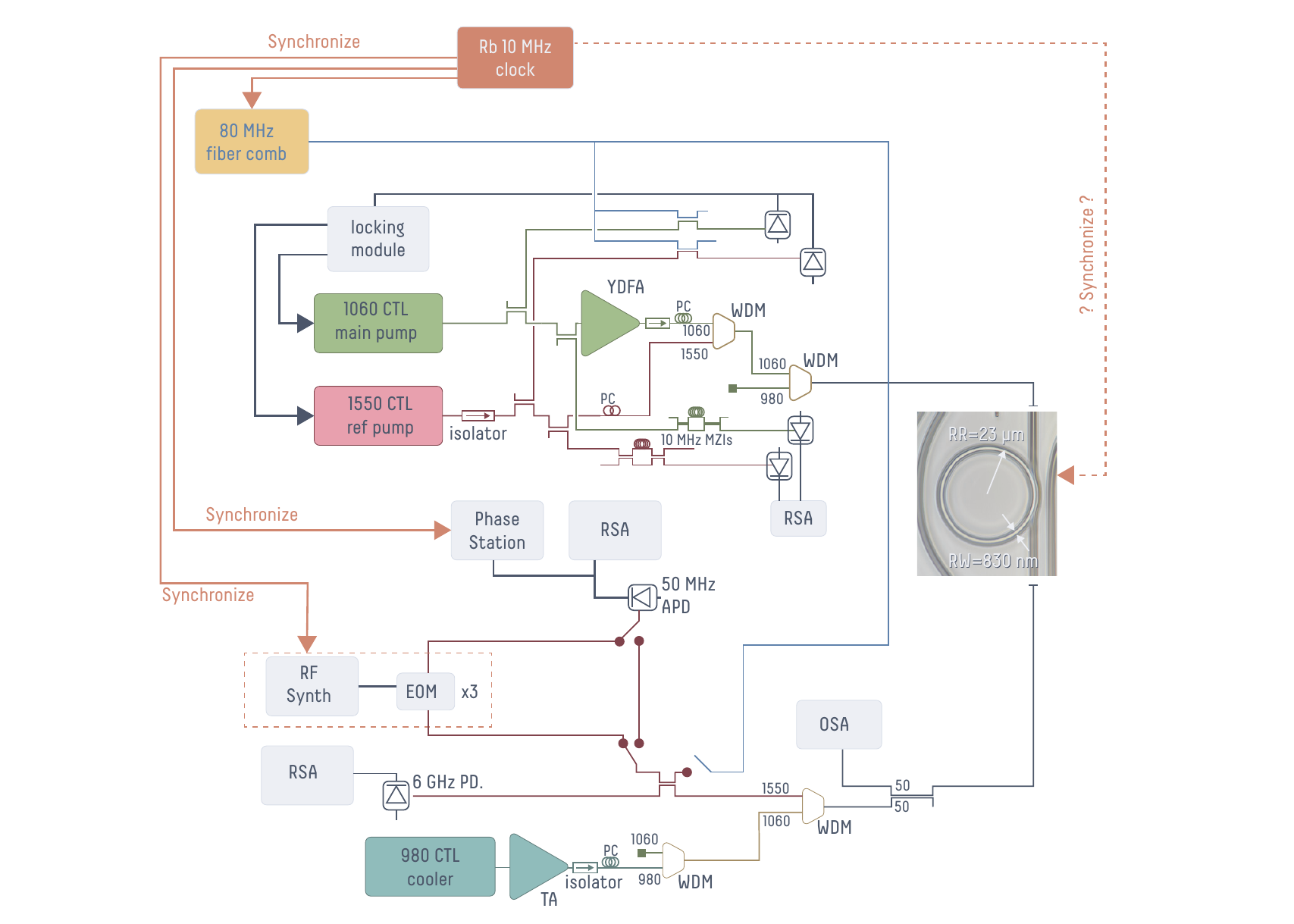}
    \end{center}
    \caption{\label{figsup:setup} Complete experimental setup. EOM: Electro-optical modulator; RSA: real time electrical spectrum analyzer; OSA: optical spectrum analyzer; WDM: wavelength demultiplexer; CTL : continuous tunable laser; MZI: Mach-Zehnder interferometer; APD: avalanche photodiode; PC: polarization controller; YDFA: Ytterbium-doped fiber amplifier; TA: tapered amplifier}
\end{figure}

The complete experimental setup is presented in \cref{figsup:setup}. Although it seems quite complex at first, most of the components are dedicated to characterizing the synchronization state, which in an application system would not have to be used. Three CTLs are used, a main pump from a 1060~nm CTL, a reference laser from a 1550~nm CTL, and a cooler pump from a 980~nm CTL. Only the reference laser is not amplified, as the synchronization does not require high power to be obtained. The polarization is set for each laser, transverse electric for the reference and main pump, and transverse magnetic for the cooler to avoid spurious nonlinear mixing in the ring. The reference and the main pump have a small amount of their power tapped for them to be locked to the 80~MHz Toptica fiber comb and sent to the MZIs for frequency noise characterization. The different lasers are combined using wavelength demultiplexers (WDMs). The cooler is filtered out after the chip to avoid high 980~nm power going upstream toward the main and reference pump. 

To characterize the DKS OFC, a portion of the light at the output of the chip is sent to the optical spectrum analyzer. The other part of the light exiting the chip is sent to different WDMs for filtering before being sent to different instruments. The CEO offset measurement presented in \cref{fig:2} is obtained by sending the 1550~nm WDM filtered light to a 6~GHz photodiode, with the output analyzed with an electrical spectrum analyzer for each reference detuning. To obtain either the repetition rate presented in \cref{fig:4} or the repetition rate noise in \cref{fig:5}, we tap a portion of the light going to the 6~GHz photodiode to either go through the EO comb apparatus or beat directly against the fiber comb. Then a narrow bandpass filter is used to select only the frequency components playing a role in the beating, which is then detected using a 50~MHz avalanche photodiode and processed either with the electrical spectrum analyzer or the phase noise analyzer.

\section{Arnold tongue bistability}
\label{sup:Arnold_bistability}

\begin{figure}[!h]
    \begin{center}
        \includegraphics{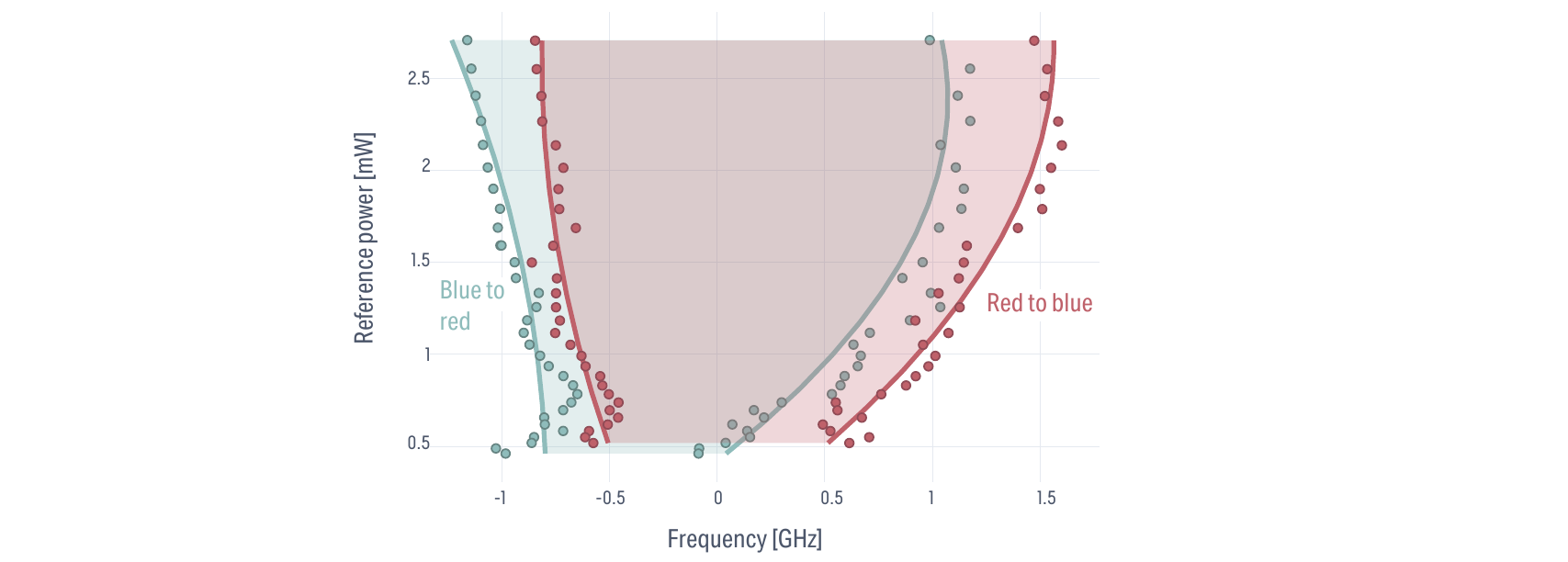}
    \end{center}
    \caption{\label{figsup:atongue} Arnold tongue hysteresis exhibiting a different behavior if probed from red to blue detuning or from blue to red. Although thermal bistability could exist, the very low reference laser power -- especially below 1~mW -- suggests that it alone is unlikely to explain the observed hysteresis. The zero is determined in both cases by a fixed frequency that is measured accurately ($<30$~MHz uncertainty) with a wavemeter.}
\end{figure}

The measurements shown in \cref{fig:2} of the synchronization bandwidth with reference power are analogous to the well-known Arnold tongue. The experiments presented in the main text are performed with reference laser detuning from red ( frequency) to blue (lower frequency). With the on-chip power of the reference below 3~mW, little thermal bistability would be expected, and no bistability at all should be expected with power below 1~mW. However, performing the Arnold tongue characterization with blue-to-red or red-to-blue tuning exhibit a hysteresis, a signature of bistability. The zero frequency of the reference detuning is the same for both scanning directions and is determined with a wavemeter with an accuracy of about 30~MHz, better than the measured hysteresis bandwidth of about 500~MHz.

\section{Frequency noise estimation from Mach-Zehnder transmission measurement}
\label{sup:mzi}

\begin{figure}[!h]
    \begin{center}
        \includegraphics[]{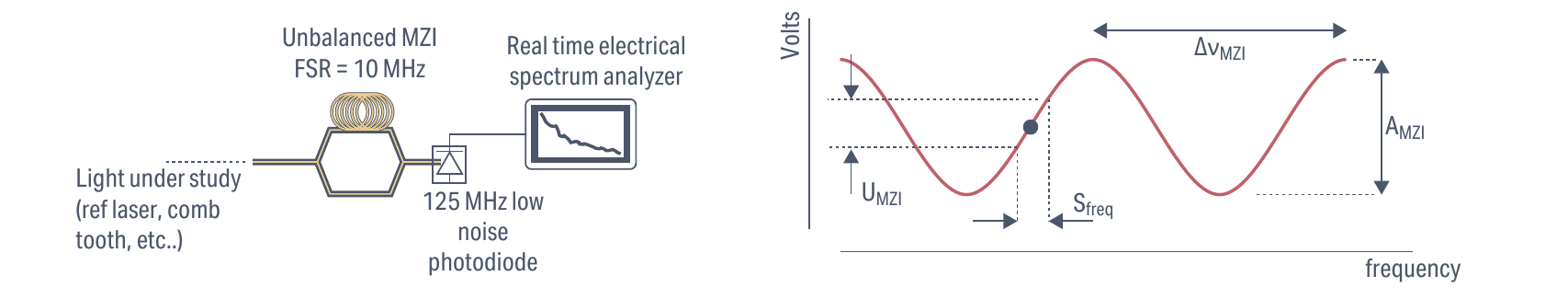}
    \end{center}
    \caption{\label{figsup:mzi}%
    \textbf{(a)} Schematic of the MZI used to measure the frequency noise of the laser. %
    \textbf{(b)} Concept to retrieve the frequency noise from the noise measurement of the light set at the quadrature point of the MZIs, converting the frequency noise into voltage noise which can be observed through an electrical spectrum analyzer.
    }
\end{figure}

In order to measure the frequency noise of the reference laser as shown in \cref{fig:5}, or of a single comb tooth as described in \cref{sup:combtooth}, we used an imbalanced fiber Mach-Zehnder interferometer with a free spectral range of 10~MHz. Working at the quadrature point allows us to convert the frequency noise of the light under study into a voltage noise which can then be observed and processed using a spectrum analyzer (RSA). To retrieve the frequency noise from the RSA, the following algebra is used:

\begin{align}
    S_\mr{\nu} = \Delta\nu_\mr{MZI} \sin^{-1}\left(\frac{U_\mr{MZI}}{A_\mr{mzi}} \right) \quad \mr{[Hz^2/Hz]}
\end{align}

\noindent with $U_\mr{MZI} = \sqrt{50 \times 10^{-3} \times 10^{S_\mr{MZI}/10}}$ the voltage obtained from spectral measurement, where $S_\mr{MZI}$ is the optical power in decibels normalized to 1~mW of the MZI noise at quadrature (i.e., in dBm), the spectrum analyzer impedance is assumed to be 50~$\Omega$, $A_\mr{mzi}$ is the amplitude of the sinusoidal modulation with frequency of the MZI, and $\Delta \nu_\mr{MZI}$ is the free spectral range of the MZI at the frequency of interest [\cref{figsup:mzi}]. 

To convert the frequency noise to phase noise, we use the following relation:
\begin{align}
    S_\varphi = 10\log_{10}\left(\frac{S_\mr{\nu}}{\nu^2}\right) \quad \mr{[dBc/Hz]}
\end{align}

\section{Phase noise and long term stability of laser locking to the fiber comb}
\label{sup:locking-noise}

\begin{figure}[!h]
    \begin{center}
        \includegraphics{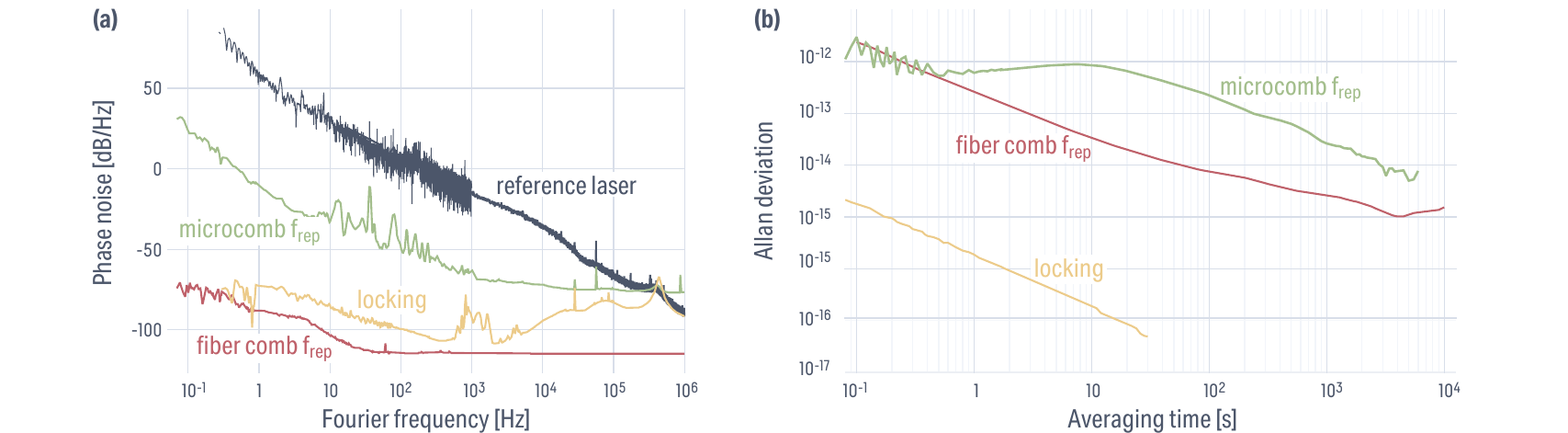}
    \end{center}
    \caption{Phase noise (a) and Allan deviation (b) of the fiber comb repetition rate (red) and laser lock (yellow) compared against the microcomb repetition rate noise (green) and reference laser noise (blue). The power in (a) is referenced to the carrier, i.e., dBc/Hz.\label{figsup:locknoise}}. 
\end{figure}

The long-term stability presented in \cref{fig:5} convincingly demonstrates that our passive synchronization architecture is compatible with an optical clockwork. We have also measured the locking phase noise and long-term stability in Fig.~\ref{figsup:locknoise} and show that, being orders of magnitude below the repetition rate of our microcomb phase noise and long-term stability, justifies their not accounted for in the final measurement presented.

    
\section{Single comb tooth noise reduction}
\label{sup:combtooth}
\begin{figure}[!h]
    \begin{center}
        \includegraphics{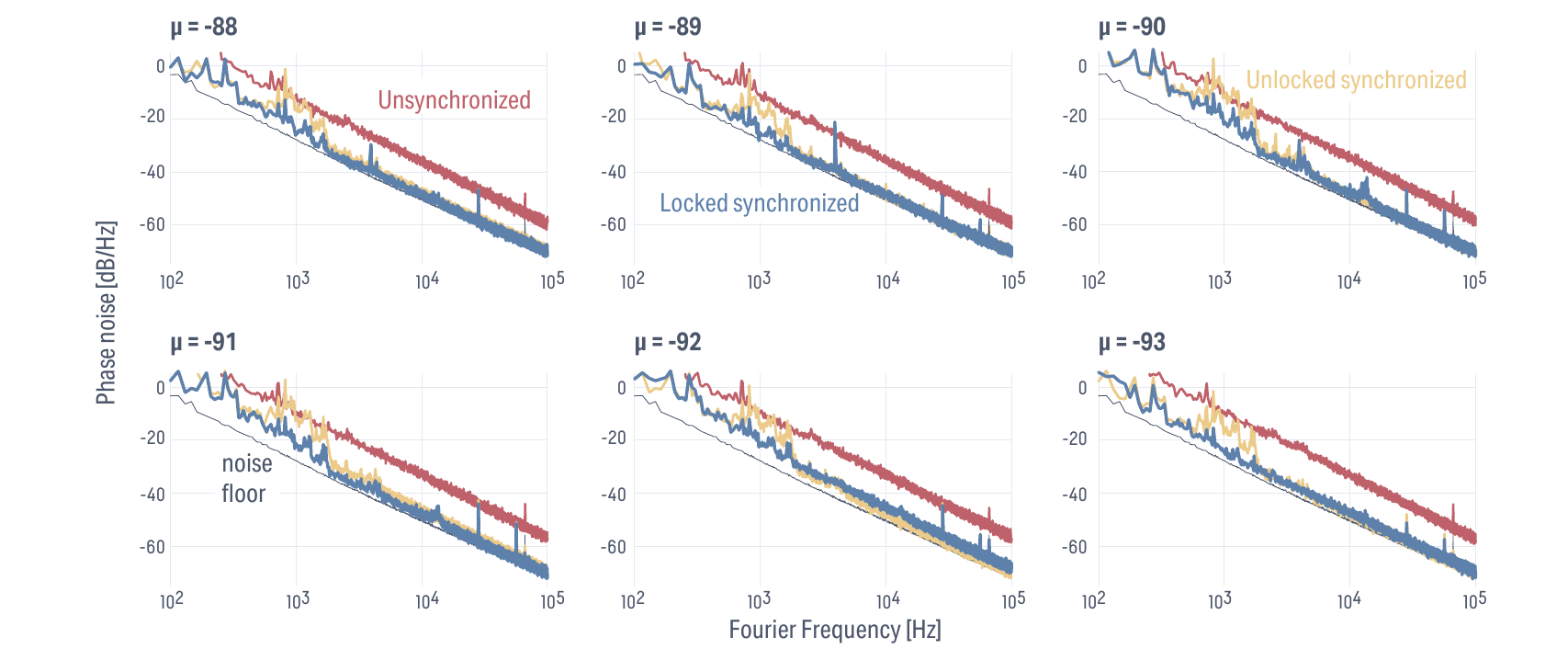}
    \end{center}
    \caption{\label{figsup:combtooth} Frequency noise of several single microcomb OFC teeth measured using the MZI apparatus from \cref{sup:mzi}, while having the reference at $\mu_s = -94$. In red, we display the unsynchronized case (\textit{i.e.} no reference present in the system), in yellow the synchronized system with all lasers being free running, and in blue the synchronized system with locked lasers. The power is referenced to the carrier, i.e., dBc/Hz.%
    }
\end{figure}

The frequency noise of the individual comb teeth of a microcomb OFC grows quadratically with the mode number, mainly since the thermo-refractive noise (TRN) is one of the dominant noise sources in such a compact cavity. However, in the synchronized case, the repetition rate is fixed by the two pinned frequencies: the reference and the CEO frequency (when the CEO is stabilized); or, in our case for the latter, the main pump frequency. To this extent, measuring the frequency noise of a single comb tooth in the different regimes is interesting. From \cref{figsup:combtooth}, it is clear that in the unsynchronized case, the noise is relatively high and follows the characteristic TRN profile. When in the synchronized case, each comb line's noise is significantly reduced and follows the noise characteristic of the pinned frequencies.

\end{document}